\documentclass[twoside,twocolumn,english,showpacs]{revtex4}
\usepackage[T1]{fontenc}
\usepackage[latin9]{inputenc}
\usepackage[active]{srcltx}
\usepackage{bm}
\usepackage{amsmath}
\usepackage{amssymb}
\usepackage{stmaryrd}
\usepackage{stackrel}
\usepackage{graphicx}

\makeatletter
\@ifundefined{textcolor}{}
{%
 \definecolor{BLACK}{gray}{0}
 \definecolor{WHITE}{gray}{1}
 \definecolor{RED}{rgb}{1,0,0}
 \definecolor{GREEN}{rgb}{0,1,0}
 \definecolor{BLUE}{rgb}{0,0,1}
 \definecolor{CYAN}{cmyk}{1,0,0,0}
 \definecolor{MAGENTA}{cmyk}{0,1,0,0}
 \definecolor{YELLOW}{cmyk}{0,0,1,0}
}

\usepackage[colorlinks,linkcolor=blue,anchorcolor=blue,citecolor=blue]{hyperref}

\makeatother

\usepackage{babel}
\begin{document}

\title{Spectral functions of a time-periodically driven Falicov-Kimball
model: Real-space Floquet dynamical mean-field theory study}

\author{Tao Qin}

\affiliation{Institut für Theoretische Physik, Goethe-Universität, 60438 Frankfurt/Main,
Germany}

\author{Walter Hofstetter}

\affiliation{Institut für Theoretische Physik, Goethe-Universität, 60438 Frankfurt/Main,
Germany}
\begin{abstract}
We present a systematic study of the spectral functions of a time-periodically
driven Falicov-Kimball Hamiltonian. In the high-frequency limit, this
system can be effectively described as a Harper-Hofstadter-Falicov-Kimball
model. Using real-space Floquet dynamical mean-field theory (DMFT),
we take into account the interaction effects and contributions from
higher Floquet bands in a non-perturbative way.
Our calculations show a high degree of similarity between the interacting
driven system and its effective static counterpart with respect to spectral
properties. However, as also illustrated by our results, one should bear in mind that Floquet DMFT describes
a non-equilibrium steady state, while an effective static Hamiltonian
describes an equilibrium state. We further demonstrate the possibility
of using real-space Floquet DMFT to study edge states on a cylinder
geometry. 
\end{abstract}
\maketitle
\section{Introduction} Time-periodically driven systems are
a versatile toolbox for simulating artificial gauge fields in experiments.
There has been great recent experimental progress in realizing topologically
nontrivial Hamiltonians. Two paradigmatic models, the Harper-Hofstadter
(HH) model~\cite{Hofstadter1976} and the Haldane model~\cite{Haldane1988},
have been realized in ultracold atoms by laser-assisted hopping~\cite{Bloch2013,Ketterle2013}
and lattice shaking~\cite{Jotzu:2014fk}, following seminal proposals~\cite{Jaksch2003,Takashi2009}.
These have triggered high interest in the physics of Floquet systems
both from a theoretical~\cite{Bukov:2014oq,Goldman:2014aa} and an
experimental~\cite{Aidelsburger:2015aa,Kennedy:2015aa} point of
view.

The effective Hamiltonian approach~\cite{Verdeny2013,Goldman:2014ab,bukov2015,Mikami2016a,Eckardt2017aqg}
works in the high-frequency limit. Different methods, such as the
Magnus expansion~\cite{bukov2015}, high-frequency expansion~\cite{Rahav:2003aa,Goldman:2014ab},
flow equations~\cite{Verdeny2013}, and Brillouin-Wigner theory~\cite{Mikami2016a},
can be used to obtain effective Hamiltonians. This approach is in
good agreement with experimental measurements~\cite{Bloch2013,Ketterle2013}
in noninteracting cases. Several efforts~\cite{Raifmmode2016ms,Bukov:2014oq,Plekhanov:2016aa}
have been made to include interaction effects within effective Hamiltonians.
Even though their validity is limited to high frequencies, they showed
the potential of Floquet engineering for obtaining exotic topologically
nontrivial phases. 

Floquet topological insulators~\cite{Kitagawa2011,Levin2013b,Carpentier2015,Rudner2015a,Eckardt2017aqg}
are a fascinating application of periodically driven systems. They
show clear differences from (static) topological insulators, especially
with respect to the bulk-edge correspondence~\cite{Levin2013b}.
Because of the periodicity of the unbounded Floquet quasienergy with
the driving frequency $\Omega$, Floquet edge states can appear in
different quasienergy gaps~\cite{Kitagawa2011,Levin2013b}. To obtain
the correct bulk-edge correspondence, in Ref.~\cite{Levin2013b} two approaches
were proposed: a winding number based on the time-evolution operator
$U(\bm{k},t)$, where $\bm{k}$ is the momentum and $t\in(0,\,\mathcal{T}]$
with the period $\mathcal{T}=\frac{2\pi}{\Omega}$, instead of $U(\bm{k},\mathcal{T})$,
and a formalism in frequency space. The former has been generalized
to the time-reversal symmetric case~\cite{Carpentier2015}. All studies
so far have focused on noninteracting models. Floquet systems are always
in a nonequilibrium state, and for the interacting case up to now
we lack a way to calculate Floquet topological invariants. The effective
topological Hamiltonian approach~\cite{Wang2012si} cannot be directly
applied to a Floquet system since one cannot easily define Green's
functions on the imaginary time axis in a nonequilibrium situation.
On the other hand, we will in the following demonstrate the strength
of real-space dynamical mean-field theory (DMFT) both for the driven system and the effective Hamiltonian,
which allows calculating edge states in the presence of interactions
$U$.

While the effective Floquet Hamiltonian describes the long-time dynamics
of Floquet systems in a nonequilibrium steady state (NESS), one needs to couple a Floquet system
to a bath, which absorbs energy from the system, in order to achieve
such a NESS~\cite{Iadecola2015b,Dehghani2014ds,Seetharam2015cb,Vorberg2013gbe}. One key issue is how to make the physical properties
of the NESS most similar to its desired effective static counterpart.
Using the quantum master or kinetic equation, Refs.~\cite{Iadecola2015b,Dehghani2014ds,Seetharam2015cb}
have studied the population of quasienergy levels in noninteracting
Floquet systems, and found that the NESS can be characterized by a
finite density of excitations above the effective Fermi level. They
have also shown how to control a NESS by different baths. The situation
is even more interesting if we introduce interactions. 

Floquet DMFT is a powerful tool for studying interacting, time-periodically
driven systems. It is a non-perturbative method for solving Hubbard-type
models with driving, and applicable in the full range of driving frequencies.
Similar to equilibrium DMFT~\cite{Georges1996}, it maps a driven,
interacting lattice model onto a single driven Anderson impurity model,
which is determined self-consistently~\cite{Freericks2008nds,Freericks2008,Aoki2014}. Every lattice
site is coupled to an additional bath in order to dissipate energy
and to help the system reach the NESS. References~\cite{Freericks2008nds,Freericks2008,Aoki2008,Aoki2009} introduced
the framework of Floquet DMFT and Refs.~\cite{Aoki2008,Aoki2009} studied the NESS of the Falicov-Kimball
model attached to a free-fermion bath and irradiated by intense light.
They calculated the electron occupation above the Fermi level due
to photon-assisted tunneling, and found a clear deviation from the
Fermi-Dirac distribution in the effective distribution. However, in
the sense of the effective Hamiltonian there is no gauge field induced
by ac driving in the model of Refs.~\cite{Aoki2008,Aoki2009}, where, instead,
only a renormalized hopping is present. In this paper, we apply Floquet
DMFT to a driven ultracold atomic system, where artificial gauge fields
are induced. We generalize the formalism of Floquet DMFT to real-space
Floquet DMFT. We find very similar spectral functions in the driven
system and its static effective counterpart when the driving frequency
is high, and a clear discrepancy for lower frequencies. We furthermore
study edge states in a cylinder geometry.

\section{The model}
We first give a short review of Flouqet's theorem. For a time-periodically
driven system, which is often termed a Floquet system, the Hamiltonian
satisfies in the time domain $H\left(t+\mathcal{T}\right)=H\left(t\right)$.
According to Floquet's theorem~\cite{bukov2015,Aoki2014}, there
is a solution of the Schr\"odinger equation of the form $\Psi_{\alpha}\left(t\right)=e^{-\mathrm{i}\varepsilon_{\alpha}t}u_{\alpha}\left(t\right)$,
where $\alpha$ labels different energy states and $u_{\alpha}\left(t\right)=u_{\alpha}\left(t+\mathcal{T}\right)$,
an analog to the Bloch function in space. $u_{\alpha}\left(t\right)$
can be Fourier expanded as $u_{\alpha}\left(t\right)=\sum_{n=-\infty}^{\infty}u_{\alpha}^{n}e^{-\mathrm{i}n\Omega t}$.
One can show that $\sum_{n}H_{mn}u_{\alpha}^{n}=\left(\varepsilon_{\alpha}+m\Omega\right)u_{\alpha}^{m}$,
where $H_{mn}=\frac{1}{\mathcal{T}}\int_{0}^{\mathcal{T}}dte^{\mathrm{i}\left(m-n\right)\Omega t}H\left(t\right)$.
$\varepsilon_{\alpha}+m\Omega$ is denoted as the quasienergy and is
not bounded. In the non-interacting case, different Floquet bands
are separated by $\Omega$. One can thus expect that the effects of higher
Floquet bands are negligible when $\Omega\rightarrow\infty$. At finite
$\Omega$, the effects from higher bands may be relevant. The situation
is more involved for interacting systems.

We now present the noninteracting Hamiltonian,
which is a fermionic version of the model realized in the experiments~\cite{Bloch2013,Ketterle2013},
\begin{align*}
H^{\left(0\right)}\left(t\right)= & H_{\mathrm{kin}}+H_{\mathrm{drive}}\left(t\right),\\
H_{\mathrm{kin}}= & -\sum_{ij}\left(J_{x}c_{i+1,j}^{\dagger}c_{ij}+J_{y}c_{i,j+1}^{\dagger}c_{ij}+H.c.\right),\\
H_{\mathrm{drive}}\left(t\right)= & \sum_{ij}\left[\frac{V_{0}}{2}\sin\left(\Omega t-\phi_{ij}+\frac{\Phi_{\boxempty}}{2}\right)+i\Omega\right]n_{ij},
\end{align*}
where $i$ and $j$ label the position $\bm{R}=i\bm{e}_{x}+j\bm{e}_{y}$
of a site in a two-dimensional square lattice, with $\bm{e}_{x}$
and $\bm{e}_{y}$ the primitive lattice vectors in the $x$ and $y$ directions.
We consider hopping terms up to nearest neighbors. $V_{0}$ is the
driving amplitude. $\Phi_{\square}$ is the flux in every primitive
unit cell, and $c_{i,j}^{\dagger}$
is the creation operator for itinerant atoms at site $\left(i,j\right)$,
with $n_{ij}=c_{i,j}^{\dagger}c_{i,j}$. For simplicity, we use the
Landau gauge $\phi_{ij}=\Phi_{\square}j$~\cite{Aidelsburger:2015aa}.
We set $J_{x}=J_{y}=1$ as our energy unit in the following.

We next derive the effective Hamiltonian. Usually, $\Omega$ is a large
energy scale in a time-periodically driven system. We perform a unitary
transformation to rotate the Hamiltonian to a frame in which there are no terms of order $\Omega$.
With the unitary rotation $\psi\left(t\right)\rightarrow V\left(t\right)\psi\left(t\right)$
where $V\left(t\right)=e^{\mathrm{i}\sum_{ij}\left(-\frac{V_{0}}{2\Omega}\cos\left(\Omega t-\phi_{ij}+\frac{\Phi_{\boxempty}}{2}\right)+i\Omega t\right)n_{ij}}$,
we have $\tilde{H}\left(t\right)=VHV^{\dagger}-\mathrm{i}V\frac{\partial V^{\dagger}}{\partial t}$.
Explicitly, 
\begin{equation}
\tilde{H}^{\left(0\right)}\left(t\right)=-\sum_{ij}\left(g\left(t\right)c_{i,j}^{\dagger}c_{i+1,j}+f\left(t\right)c_{i,j}^{\dagger}c_{i,j+1}+h.c.\right),\label{eq:rotH}
\end{equation}
where $g\left(t\right)=e^{\mathrm{i}\mathcal{A}\sin\left(\Omega t-\phi_{ij}\right)-\mathrm{i}\Omega t}$
and $f\left(t\right)=e^{\mathrm{i}\mathcal{A}\sin\left(\Omega t-\phi_{ij}\right)}$,
and we define $\mathcal{A}\equiv\frac{V_{0}}{\Omega}\sin\frac{\Phi_{\boxempty}}{2}\equiv\frac{V_{0}}{\Omega}\sin\left(\pi\alpha\right)$
with $\Phi_{\square}=2\pi\alpha$. Using the Magnus expansion~\cite{bukov2015},
the effective Hamiltonian up to zeroth order in $\frac{1}{\Omega}$
is 
\begin{align}
\tilde{H}_{\mathrm{eff}}^{\left(0\right)} & =\frac{1}{\mathcal{T}}\int_{0}^{\mathcal{T}}dt\tilde{H}^{\left(0\right)}\left(t\right)=-\sum_{i,j}\left[\mathcal{J}_{1}\left(\mathcal{A}\right)e^{-\mathrm{i}\phi_{ij}}c_{i,j}^{\dagger}c_{i+1,j}\right.\nonumber \\
 & \left.+\mathcal{J}_{0}\left(\mathcal{A}\right)c_{i,j}^{\dagger}c_{i,j+1}+H.c.\right]\label{eq:effH}
\end{align}
where $\mathcal{J}_{l}\left(\mathcal{A}\right)$ is the $l$-th order
Bessel function of the first kind. This effective Hamiltonian is exact
in the limit $\Omega\rightarrow\infty$ while higher orders contribute
for finite $\Omega$. The Hamiltonian~(\ref{eq:effH}) is slightly
different from the usual HH model, since the hopping amplitudes $\mathcal{J}_{l}\left(\mathcal{A}\right)$
depend on the flux $\alpha$ through $\mathcal{A}$. 

We are now in a position to present Hamiltonian~(\ref{eq:rotH})
in Floquet space. Using Floquet's theorem~\cite{Gomezleon2013} and
an extended Hilbert space~\cite{Sambe1973}, we can write $c_{i,j}\left(t\right)=\sum_{n=-\infty}^{\infty}c_{i,j,n}e^{-\mathrm{i}n\Omega t}$.
The Hamiltonian $\tilde{H}^{\left(0\right)}\left(t\right)$ in the
Heisenberg picture can be transformed to Floquet space. Its matrix
element $\mathcal{H}_{mn}^{\left(0\right)}$ is given by 
\begin{align}
\mathcal{H}_{mn}^{\left(0\right)} & =-\sum_{i,j}\left[\mathcal{J}_{n-m+1}\left(\mathcal{A}\right)e^{\mathrm{i}\left(m-n-1\right)\phi_{ij}}c_{i,j,m}^{\dagger}c_{i+1,j,n}\right.\nonumber \\
 & +\mathcal{J}_{m-n+1}\left(\mathcal{A}\right)e^{\mathrm{i}\left(m-n+1\right)\phi_{ij}}c_{i+1,j,m}^{\dagger}c_{i,j,n}\nonumber \\
 & +\mathcal{J}_{n-m}\left(\mathcal{A}\right)e^{\mathrm{i}\left(m-n\right)\phi_{ij}}c_{i,j,m}^{\dagger}c_{i,j+1,n}\nonumber \\
 & \left.+\mathcal{J}_{m-n}\left(\mathcal{A}\right)e^{\mathrm{i}\left(m-n\right)\phi_{ij}}c_{i,j+1,m}^{\dagger}c_{i,j,n}\right],\label{eq:floquetH}
\end{align}
which in physical terms corresponds to the stimulated emission ($m>n$)
or absorption ($m<n$) of photons~\cite{Aoki2008}. The diagonal
terms correspond to the strongest hopping, while the off-diagonal parts
are higher-order corrections. In principle, the dimensionality of
$\mathcal{H}^{\left(0\right)}$ is infinite. In reality, we can keep
a finite matrix, with a size inversely proportional to the driving
frequency $\Omega$. We also make the important observation that Hamiltonian~(\ref{eq:floquetH})
recovers the effective Hamiltonian~(\ref{eq:effH}) if we set the
dimension of Floquet space to 1. 

\section{Floquet DMFT formalism and its real-space generalization} We
are mostly interested in the interaction effects in the driven system.
We turn on an interaction of the Falicov-Kimball type,
\begin{equation}
H_{\mathrm{int}}=U\sum_{i,j}c_{i,j}^{\dagger}c_{i,j}f_{i,j}^{\dagger}f_{i,j},\label{eq:FK}
\end{equation}
where $f_{i,j}^{\dagger}$ is the creation operator of a particle
of the localized species at site $\left(i,j\right)$. 

To gain comprehensive understanding, our calculation should (i) take
into account the effect of the interaction, and (ii) the contribution
arising from higher orders in $\frac{1}{\Omega}$. While the former
can be achieved by equilibrium DMFT based on the effective Hamiltonian~\cite{Georges1996},
both (i) and (ii) can be achieved by Floquet DMFT~\cite{Freericks2008nds,Freericks2008,Aoki2008,Aoki2009,Tsuji2010,Lubatsch2009,Lee2014di,Aoki2014}.
In this work, we go one step further by generalizing Floquet DMFT
calculations to inhomogeneous systems~(see the Appendix Sec.~\ref{subsec:real-FloquetRDMFT}). 

\section{Hofstadter butterfly} 
\begin{figure}[h]
\includegraphics[scale=0.7]{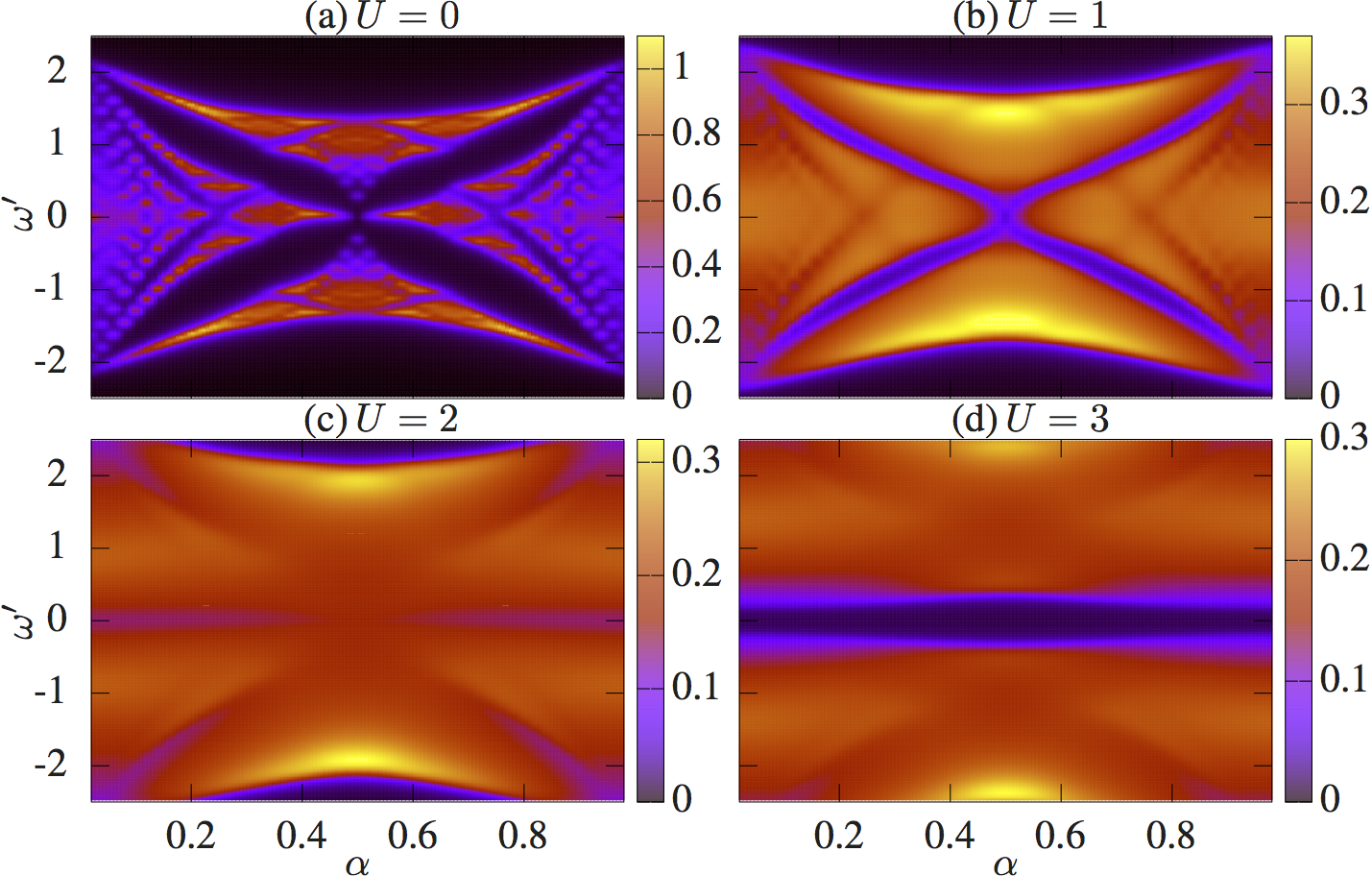}

\caption{\label{fig:butterfly}Hofstadter butterfly: Spectral function $A_{ij}\left(\omega^{\prime}\right)=-\frac{1}{\pi}\mathrm{Im}G_{ij,nn}^{R}\left(\omega\right)$
for the center site of a 15$\times$15 square lattice, calculated
by real-space Floquet DMFT. }
\end{figure}

We first present a real-space
Floquet DMFT calculation for the local spectral function $A_{ij}\left(\omega^{\prime}\right)=-\frac{1}{\pi}\mathrm{Im}G_{ij,nn}^{R}\left(\omega\right)$
on site $\left(i,j\right)$ of a driven Falicov-Kimball model at half filling
with $w_{0}=w_{1}=\frac{1}{2}$. $G_{ij,nn}^{R}\left(\omega\right)$
is the $\left(n,n\right)$ Floquet component of the nonequilibrium
retarded Green's function on site $\left(i,j\right)$~(see the Appendix Sec.~\ref{subsec:real-FloquetRDMFT}).
$\omega\in\left(-\frac{\Omega}{2},\frac{\Omega}{2}\right]$, and $\omega^{\prime}\equiv\omega+n\Omega$
is in the full range of the frequency spectrum. The spectral function is
shown for the center site of a 15$\times$15 square lattice, which
is in the bulk and preserves the symmetry of the system. We have chosen
values for the interaction $U$ which are significantly lower than
the driving frequency $\Omega=7$. It is therefore expected that the
effective HH Hamiltonian can capture the features of the driven Falicov-Kimball
model. Indeed, in Fig.~\ref{fig:butterfly}, we clearly observe a
Hofstadter butterfly structure. Furthermore, we see that increasing
interaction smears out the fine structure of the butterfly, consistent
with DMFT calculations for the HH-Falicov-Kimball model~\cite{Tran2010}.
These results clearly demonstrate the possibility to observe the Hofstadter
butterfly in a Floquet system. 

We next have a close look at the properties of spectral functions for
typical fluxes $\alpha=\frac{1}{6}$ and $\alpha=\frac{1}{4}$, where
the latter is already realized in experiments~\cite{Bloch2013,Aidelsburger:2015aa}.
To this end, we compare results from real-space DMFT calculations
for an effective HH-Falicov-Kimball model, and real-space Floquet
DMFT calculations for a driven Falicov-Kimball model, which we refer
to, respectively, as the static and driven cases from now on. 

\begin{figure}[h]
\includegraphics[scale=0.7]{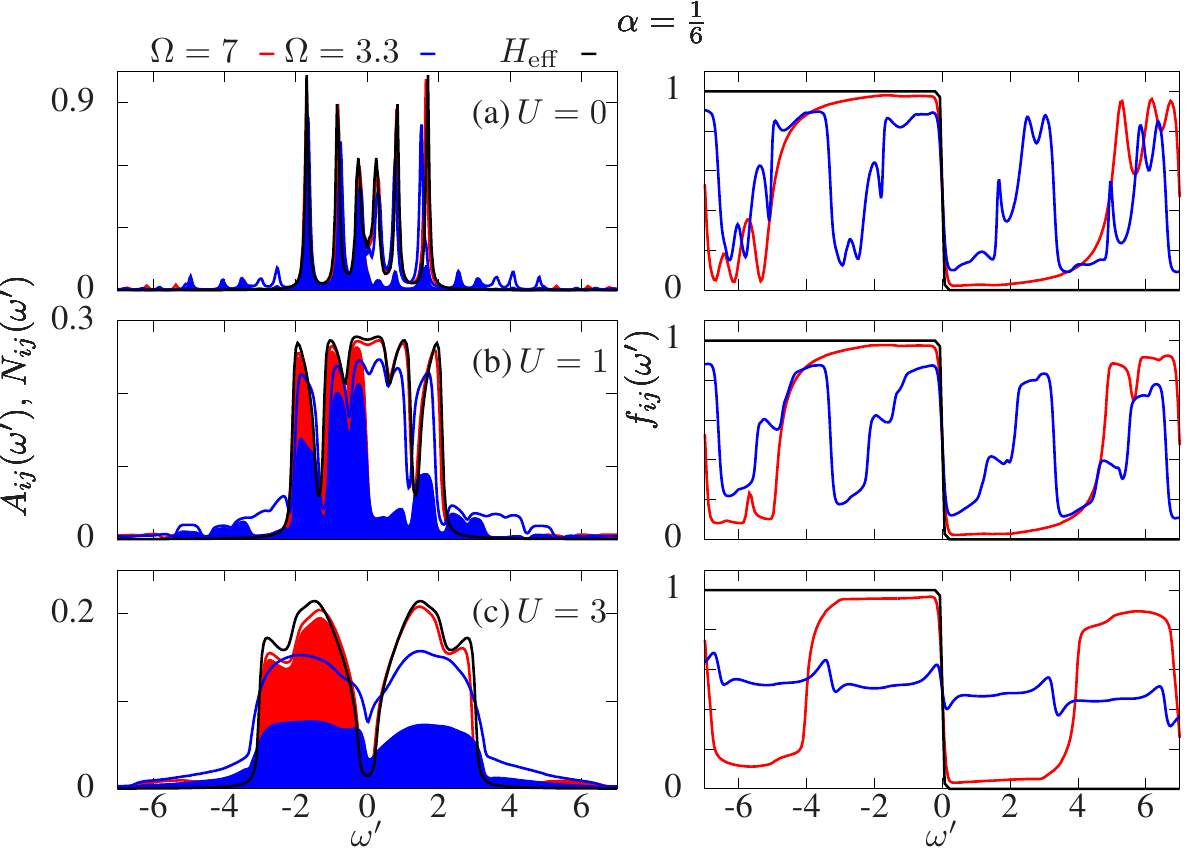}
\includegraphics[scale=0.7]{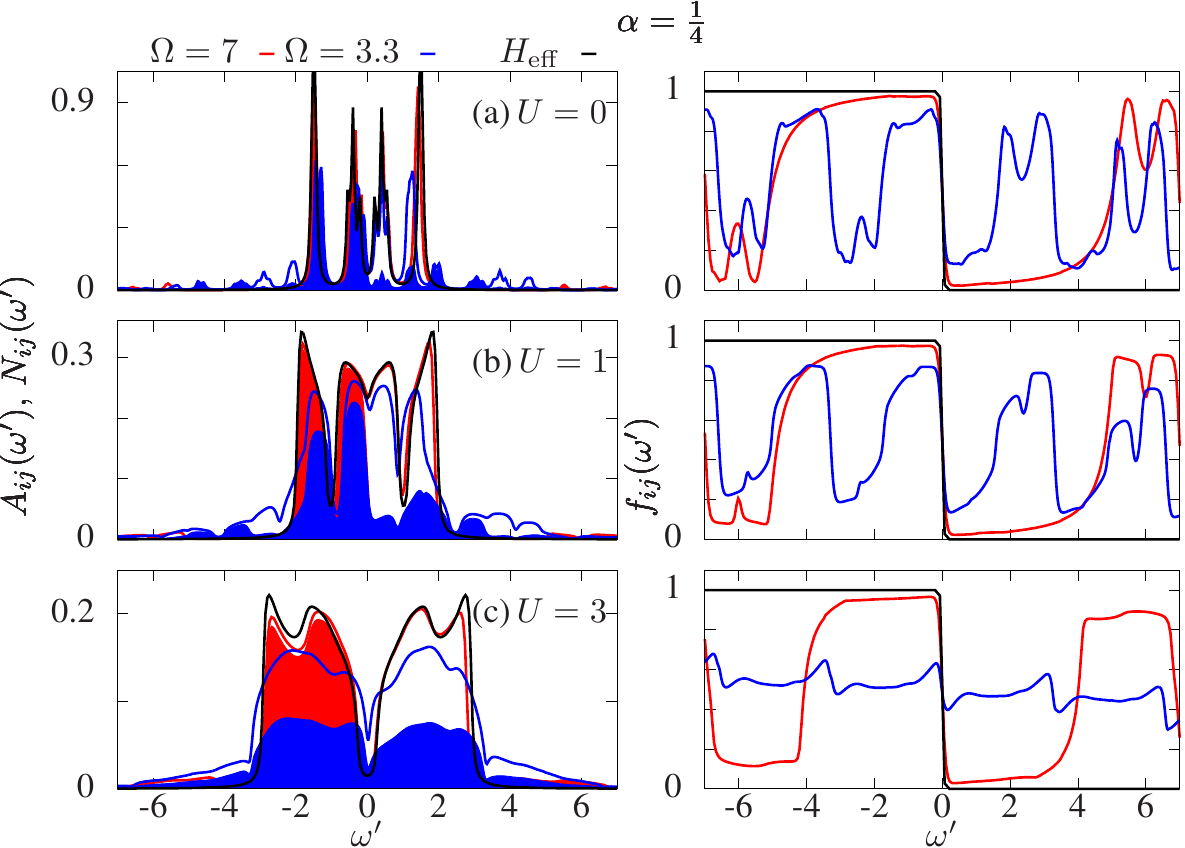}
\caption{\label{fig:1o6}Spectral function $A_{ij}\left(\omega^{\prime}\right)=-\frac{1}{\pi}\mathrm{Im}G_{ij,nn}^{R}\left(\omega\right)$
(lines in the left panel), occupied density of states $N_{ij}\left(\omega^{\prime}\right)=\frac{1}{2\pi}\mathrm{Im}G_{ij,nn}^{<}\left(\omega\right)$
(shaded areas in the left panel for driven cases), and effective distribution
$f_{ij}\left(\omega^{\prime}\right)=N_{ij}\left(\omega^{\prime}\right)/A_{ij}\left(\omega^{\prime}\right)$
(lines in the right panel) at the center site of a $15\times15$ square
lattice with $\alpha=\frac{1}{6}$ (upper panel) and $\alpha=\frac{1}{4}$
(lower panel), for the driven case (frequencies $\Omega=3.3$ and
$\Omega=7$) and the static effective Hamiltonian.}
\end{figure}

In Fig.~\ref{fig:1o6}, we show the static and driven spectral functions
for different fluxes and driving frequencies. (i) We observe six peaks
in the spectrum for $\alpha=\frac{1}{6}$ and four peaks for $\alpha=\frac{1}{4}$,
when $U=0$. With increasing interaction $U$, the peaks decay, and
the fine structure of the butterfly is smeared out. Only two Mott-insulator
bands are left for $U=3$. (ii) We observe good consistency of the
spectral functions in the driven case at high frequency and the static
case. This similarity is very important and justifies the use of the
NESS for simulating the equilibrium state. On the other hand, we see
a clear discrepancy between the static and driven cases at an intermediate
driving frequency $\Omega=3.3$. (iii) Note that the spectral functions
for the driven system are not as symmetric as those for the static
case. This is due to contributions from higher Floquet bands, since
the Floquet Hamiltonian reduces to the effective HH-Falicov-Kimball
Hamiltonian if we set the dimension of the Floquet matrix to 1. (iv)
The static and driven cases are essentially different, which is shown
by the effective distribution $f_{ij}\left(\omega^{\prime}\right)=N_{ij}\left(\omega^{\prime}\right)/A_{ij}\left(\omega^{\prime}\right)$
with $N_{ij}\left(\omega^{\prime}\right)=\frac{1}{2\pi}\mathrm{Im}G_{ij,nn}^{<}\left(\omega\right)$
in Fig.~\ref{fig:1o6}. While for the static case the distribution
is naturally of a Fermi-Dirac type, it is clearly different in the NESS
of the driven system. The effective distribution thus qualitatively
describes how far the NESS is from an equilibrium state. 

\begin{figure}[h]
\includegraphics{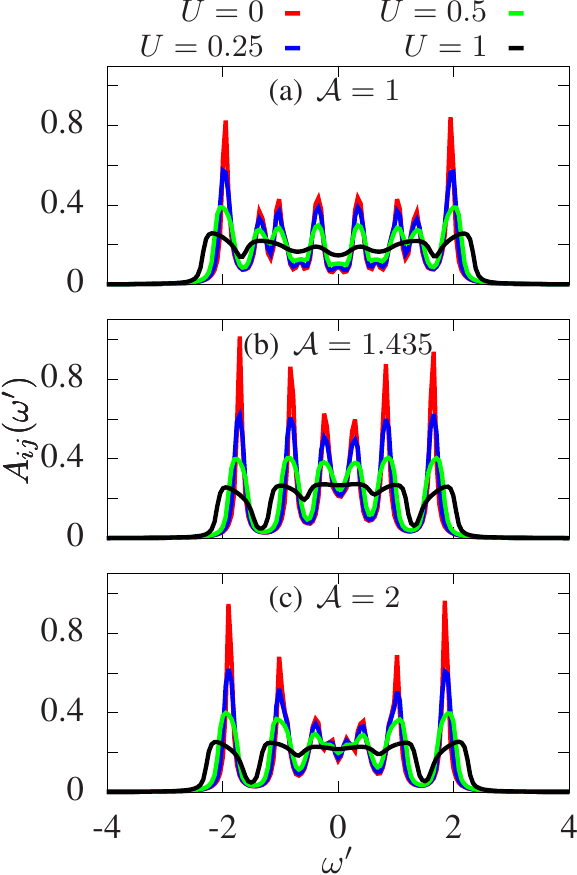}

\caption{\label{fig:amplitudes}Spectral function $A_{ij}\left(\omega^{\prime}\right)=-\frac{1}{\pi}\mathrm{Im}G_{ij,nn}^{R}\left(\omega\right)$
at the center site of a $15\times15$ square lattice with different
driving amplitudes for $\Omega=7$ and $\alpha=\frac{1}{6}$. }
\end{figure}

We need to point out that in these plots of spectral functions we
choose the amplitude $V_{0}$ to satisfy the condition $\mathcal{A}=1.435$,
i.e., $V_{0}=1.435\frac{\Omega}{\sin\left(\alpha\pi\right)}$, where
$\mathcal{J}_{0}\left(\mathcal{A}\right)=\mathcal{J}_{1}\left(\mathcal{A}\right)$.
In this way, the effective hopping amplitudes in $x$ and $y$ are
the same~\cite{experimentnote}, which makes the effective Hamiltonian exactly equivalent
to the standard HH model. To illustrate this point, in Fig.~\ref{fig:amplitudes}
we show the difference between $\mathcal{A}=1$, $1.435$, and $2$
for $\Omega=7$ and $\alpha=\frac{1}{6}$. The choices $\mathcal{A}=1$
and $2$ lead to additional peaks in the spectrum. It is therefore
of experimental relevance to choose the special driving amplitude
satisfying $\mathcal{A}=1.435$. To observe a Hofstadter butterfly,
the temperature should be smaller than the gaps between subbands~\cite{Chin2014}.
Additional peaks for a nonoptimal choice of $\mathcal{A}$ would
imply even smaller gaps and increase the experimental challenge. 

\begin{figure}[h]
\includegraphics[scale=0.73]{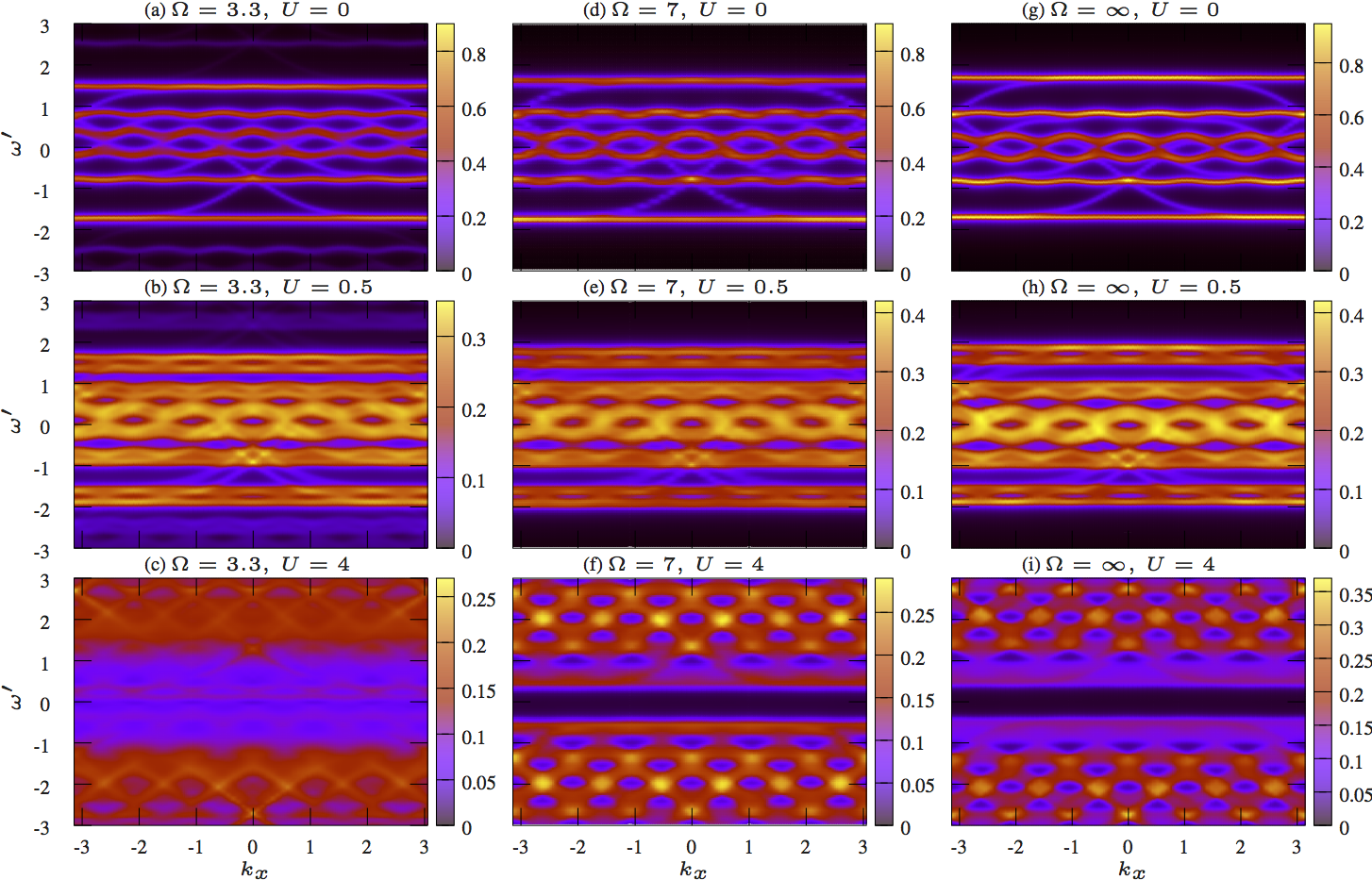}\caption{\label{fig:edge}Total spectrum $A\left(k_{x},\omega^{\prime}\right)=\frac{1}{N_{y}}\sum_{\bm{R}_{y}}A\left(k_{x},\bm{R}_{y},\omega^{\prime}\right)$
with $N_{y}=35$, including edge and bulk states for the driven and effective
static ($\Omega=\infty$) cases on a cylinder geometry, which is periodic
in the $x$ direction and finite in the $y$ direction, with $\alpha=\frac{1}{6}$.
For the driven cases, $V_{0}=1.435\frac{\Omega}{\sin\left(\alpha\pi\right)}$.}
\end{figure}
\section{Edge state on a cylinder structure} In Fig.~\ref{fig:edge}
we present spectral functions for the static and driven cases in a
cylinder geometry, which is periodic in the $x$ direction and finite
in the $y$ direction, using real-space (Floquet) DMFT. We show the total
spectrum containing bulk and edge states. We compare properties of
the driven system for different driving frequencies with the static
case at the same $U$. The driven case at high frequency ($\Omega=7$)
is very similar to the static one, even though there is a slight difference
in the symmetry of the spectrum, which is, again, due to contributions
from higher Floquet bands. For intermediate frequency $\Omega=3.3$,
higher Floquet bands become visible at $\omega\approx\pm2.5$. With
increasing $U$, we observe that fine structures are washed out, and
finally a gap opens. For $U=4$ and $\Omega=3.3$, we observe a clear
difference compared to the high-frequency case. It may be due to photon-assisted
tunneling inside the Floquet band around $\omega=0$. When $U=0$,
edge states are present for all driving frequencies. Overall, we see
a pronounced similarity between $\Omega=7$ and the effective static
case ($\Omega=\infty$). This justifies the idea to engineer nontrivial
topological states by driving. Because the spectrum is washed out
with increasing interactions, edge states become invisible at large
$U$. Even though we do not observe edge states \emph{induced} by
strong interactions in this simple model, it may be possible to observe
this effect in other models. We emphasize that it is crucial to study edge states
in an interacting Floquet system, since currently there is no approach
for calculating the Chern number or related topological indices of
interacting non-equilibrium states. Therefore, edge states provide
important signatures for topological transitions.

\section{Experimental realization} The set up and results
we have presented are accessible for experimental measurements. The
spectral functions presented above can be detected using momentum-resolved
radio-frequency (rf) spectroscopy~\cite{Torma2016ps,Stewart:2008aa},
a counterpart to angle-resolved photoemission spectroscopy (ARPES)
used for electronic materials. We have some remarks on the bath. The
role of the free-fermion bath is to allow a driven system to reach
a NESS. It has been proposed~\cite{Griessner2006di} that atoms in
an optical lattice can be cooled by immersion in a Bose-Einstein condensate
(BEC), which also serves as a bath. 

Reference~\cite{Jotzu2015a} demonstrates a novel approach, which has
the potential to realize a Falicov-Kimball model, where different
species (hyperfine states) can be subjected to different driving amplitudes.
For a system of two species, one species can be ``dynamically localized''
by tuning the renormalized hopping amplitude to zero. Therefore, experimental
techniques for realizing both the Falicov-Kimball model and the HH
model are available. 

\section{Conclusion}  Time-periodically driven ultracold atoms
are a promising platform for simulating topologically nontrivial band
structures. In the presence of interactions, these are even more intriguing
and interesting. Using real-space (Floquet) DMFT, we have systematically
studied the spectral function of the driven Falicov-Kimball Hamiltonian,
and of its effective Hamiltonian in the high-frequency limit, both
for open boundary conditions and for a cylinder geometry. For a large
driving frequency we observed similar spectra of the driven system and the effective Hamiltonian. This demonstrates that
topologically nontrivial bands can be simulated by a realistic driven
system. Nevertheless, as we have shown, the NESS of the driven Hamiltonian
is essentially different from the equilibrium state of an effective
Hamiltonian. Our work also highlights the possibility of studying
edge states and topological properties of an interacting system by using real-space
Floquet DMFT.

\section{Acknowledgments}
The authors acknowledge useful discussions and communication with
M. Eckstein, K. Le Hur, and N. Tsuji. This work was supported by the
Deutsche Forschungsgemeinschaft via DFG FOR 2414 and the high-performance
computing center LOEWE-CSC.


\section{Appendix}
\subsection{\label{subsec:real-FloquetRDMFT} Floquet DMFT and its real-space generalization}
\begin{figure*}[t]
\centering
\includegraphics[width=0.7\textwidth]{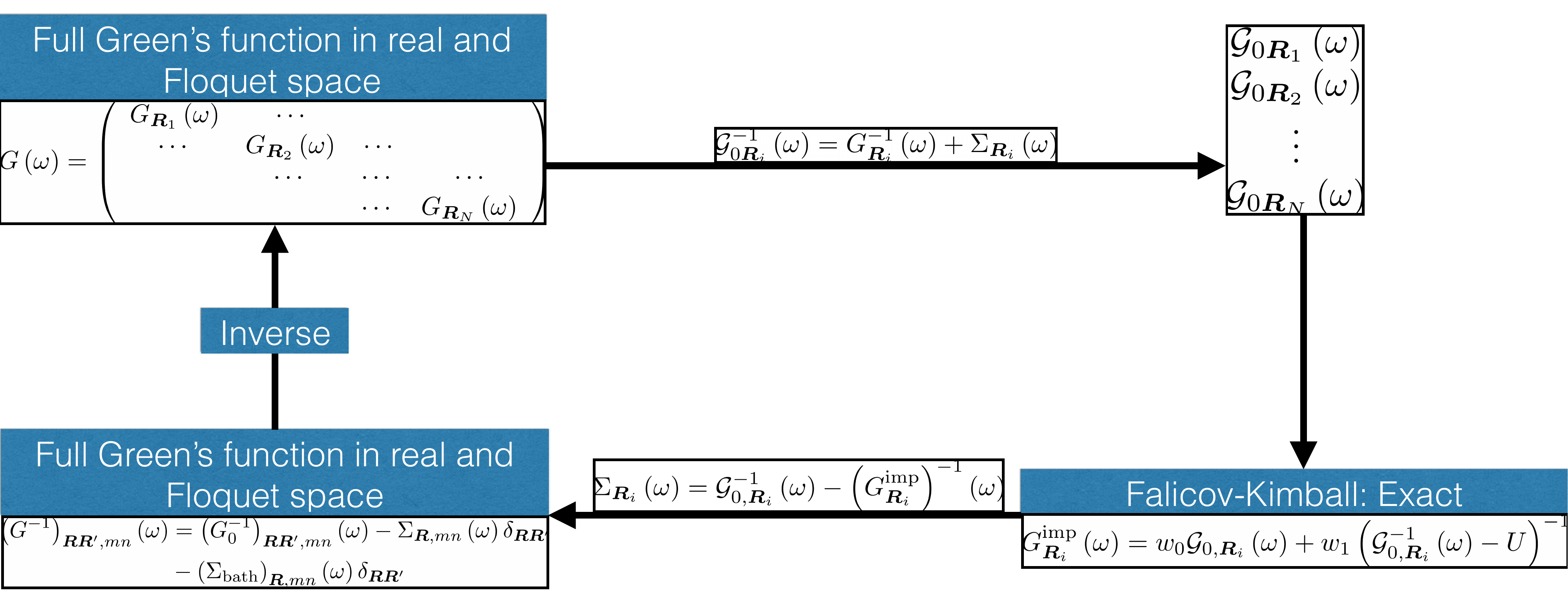}\caption{\label{fig:dmft}Flow chart of real-space Floquet DMFT. All Green's
functions are defined on the Keldysh contour and in real and Floquet
space. }
\end{figure*}
While the equilibrium real-space DMFT formalism has been detailed
in Ref.~\cite{Snoek2008}, here we give a short introduction to Floquet
DMFT and its real-space generalization. We first highlight two important
aspects. (i) Floquet DMFT addresses the NESS in a non-equilibrium
system~\cite{Freericks2008nds,Freericks2008,Aoki2014}, and is based on the Keldysh Floquet Green's
function~\cite{Aoki2008,Tsuji2010}. Because of driving, we need
to consider the two-time Green's function $G\left(t,t^{\prime}\right)=G\left(t+\mathcal{T},t^{\prime}+\mathcal{T}\right)\neq G\left(t-t^{\prime}\right)$,
where $t$ and $t^{\prime}$ are defined on a two-branch contour,
ranging from $-\infty$ to $+\infty$, and from $+\infty$ to $-\infty$,
respectively. The Green's function in frequency space is $G_{n}\left(\omega\right)=\int_{-\infty}^{\infty}dt_{\mathrm{rel}}\frac{1}{\mathcal{T}}\int_{0}^{\mathcal{T}}dt_{\mathrm{av}}G\left(t_{\mathrm{rel}},t_{\mathrm{av}}\right)e^{\mathrm{i}\omega t_{\mathrm{rel}}+\mathrm{i}n\Omega t_{\mathrm{av}}}$,
where $t_{\mathrm{rel}}=t-t^{\prime}$ and $t_{\mathrm{av}}=\frac{t+t^{\prime}}{2}$.
Generally, one can calculate the noninteracting Green's function
analytically and obtain the interacting Green's function using the
Dyson equation. To keep a structure of the Dyson equation that is
convenient for calculations, one needs to use the Floquet Green's
function. It is defined by the map~\cite{Aoki2008} $G_{mn}\left(\omega\right)=G_{m-n}\left(\omega+\frac{m+n}{2}\Omega\right)$,
where $\omega\in\left(-\frac{\Omega}{2},\frac{\Omega}{2}\right]$.
(ii) To achieve a NESS, every lattice site is coupled to a bath, which
extracts energy from the driven lattice. The bath can be fermionic~\cite{Aoki2009,Tsuji2010,Aoki2014}
or bosonic~\cite{Lee2014di}. We consider a free-fermion bath in
our calculations. The effect of the bath is equivalent to a correction
to the self-energy, namely~\cite{Aoki2009,Tsuji2010,Aoki2014}, 
\begin{equation}
G_{\bm{k}}^{-1}\left(\omega\right)=G_{\bm{k}0}^{-1}\left(\omega\right)-\Sigma\left(\omega\right)-\Sigma_{\mathrm{bath}}\left(\omega\right).\label{eq:FloquetDMFT}
\end{equation}
All quantities here have three components, for example, $G_{\bm{k}}\left(\omega\right)=\left(\begin{array}{cc}
G_{\bm{k}}^{R}\left(\omega\right) & G_{\bm{k}}^{K}\left(\omega\right)\\
0 & G_{\bm{k}}^{A}\left(\omega\right)
\end{array}\right)$, and every component is a Floquet matrix. Within a flat density of
states approximation~\cite{Aoki2014,Tsuji2010}, $\Sigma_{\mathrm{bath}}\left(\omega\right)=\left(\begin{array}{cc}
\mathrm{i}\Gamma\hat{\mathrm{I}} & -2\mathrm{i}\Gamma F\left(\omega\right)\hat{\mathrm{I}}\\
0 & -\mathrm{i}\Gamma\hat{\mathrm{I}}
\end{array}\right)$ with the $n$ th component of $F\left(\omega\right)$ given by  $F_{n}\left(\omega\right)=\tanh\frac{\hbar\left(\omega+n\Omega\right)}{2k_{B}T}$,
where the bath is described by two parameters: damping rate $\Gamma$
and bath temperature $T$. $\hat{\mathrm{I}}$ is the unit matrix.
The noninteracting part is $G_{\bm{k}0}^{-1}\left(\omega\right)=\left(\begin{array}{cc}
\left(G_{\bm{k}0}^{R}\right)^{-1}\left(\omega\right) & \left(G_{\bm{k}0}^{-1}\left(\omega\right)\right)^{K}\\
0 & \left(G_{\bm{k}0}^{A}\right)^{-1}\left(\omega\right)
\end{array}\right)$, where $\left(G_{\bm{k}0}^{R\left(A\right)}\right)^{-1}\left(\omega\right)$
can be determined from the noninteracting Hamiltonian, and it can
be shown from the fluctuation-dissipation theorem that $\left(G_{\bm{k}0}^{-1}\left(\omega\right)\right)^{K}$
is negligible~\cite{Aoki2009,Aoki2014}. For a driven Hubbard model
we can use iterative perturbation theory~\cite{Georges1996,Aoki2014}
as an impurity solver. For the driven Falicov-Kimball model the effective
impurity problem can be solved analytically in infinite dimensions.

We next describe the formalism for generalizing Floquet DMFT to a
position-dependent self-energy, which is suitable for studying an
inhomogeneous system. We have a lattice of driven, effective impurity
models, which are coupled via the lattice Dyson equation, which is
the real-space version of Eq.~(\ref{eq:FloquetDMFT}), 
\begin{align}
\left(G^{-1}\right)_{\bm{R}\bm{R}^{\prime},mn}\left(\omega\right) & =\left(G_{0}^{-1}\right)_{\bm{R}\bm{R}^{\prime},mn}\left(\omega\right)-\Sigma_{\bm{R},mn}\left(\omega\right)\delta_{\bm{R}\bm{R}^{\prime}}\nonumber \\
 & -\left(\Sigma_{\mathrm{bath}}\right)_{\bm{R},mn}\left(\omega\right)\delta_{\bm{R}\bm{R}^{\prime}}.\label{eq:realdyson}
\end{align}
The noninteracting Floquet Green's function is $\left(G_{0}^{-1}\right)_{\bm{R}\bm{R}^{\prime},mn}^{R\left(A\right)}\left(\omega\right)=\left(G_{0}^{R\left(A\right)}\right)_{\bm{R}\bm{R}^{\prime},mn}^{-1}\left(\omega\right)=\omega+n\Omega-\mathcal{H}_{\bm{R}\bm{R}^{\prime},mn}^{\left(0\right)}\pm\mathrm{i}0^{+}$,
where $\mathcal{H}_{\bm{R}\bm{R}^{\prime},mn}^{\left(0\right)}$ is
given by Eq.~\eqref{eq:floquetH} in the main text, with lattice sites $\bm{R}=i\bm{e}_{x}+j\bm{e}_{y}$
and $\bm{R}^{\prime}=i^{\prime}\bm{e}_{x}+j^{\prime}\bm{e}_{y}$.
$i$ ($i^{\prime}$) and $j$ ($j^{\prime}$) label the $x$ and $y$
coordinates of the sites of a finite lattice, and $m$ and $n$ are
indices of a Floquet matrix. $\left(G_{0}^{-1}\right)_{\bm{R}\bm{R}^{\prime},mn}^{K}\left(\omega\right)$
is again negligible. The self-energies $\Sigma_{\bm{R}}$ and $\Sigma_{\mathrm{bath},\bm{R}}$
are diagonal in position space, and $\Sigma_{\bm{R}}$ is determined self-consistently
by an impurity solver. 

For the interaction in Eq.~\eqref{eq:FK} in the main text, we adopt the following
solver for each one of the effective impurity problems, i.e., for the
impurity at site $\bm{R}$ the Floquet Green's function is
\begin{equation}
G_{\bm{R}}\left(\omega\right)=w_{0}\mathcal{G}_{0,\bm{R}}\left(\omega\right)+w_{1}\left[\mathcal{G}_{0,\bm{R}}^{-1}\left(\omega\right]-U\right)^{-1},\label{eq:realsolver}
\end{equation}
where $\mathcal{G}_{0,\bm{R}}\left(\omega\right)$ is the Weiss function
and determined self-consistently, $w_{1}$ is the filling of the localized
$f$ atoms, and $w_{0}=1-w_{1}$. Equations~(\ref{eq:realdyson}),
(\ref{eq:realsolver}), and the impurity Dyson equation 
\begin{equation}
\mathcal{G}_{0,\bm{R}}^{-1}\left(\omega\right)=G_{\bm{R}}^{-1}\left(\omega\right)+\Sigma_{\bm{R}}\left(\omega\right)
\end{equation}
form the set of self-consistency equations of real-space Floquet DMFT (Fig.~\ref{fig:dmft}).

\subsection{Bath effects}

\begin{figure}[h]
\includegraphics{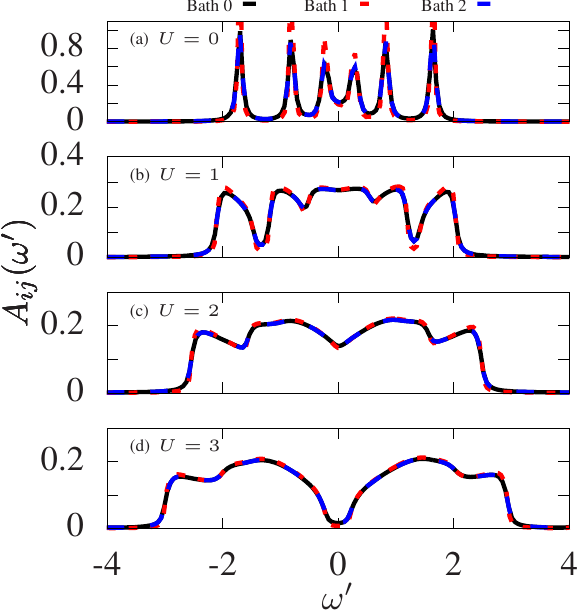}

\caption{\label{fig:bath}Spectral functions $A_{ij}\left(\omega+n\Omega\right)=-\frac{1}{\pi}\mathrm{Im}G_{ij,nn}^{R}\left(\omega\right)$
at the center site of a $15\times15$ square lattice with different
baths for $\Omega=7$ and $\alpha=\frac{1}{6}$. ``Bath 0'' with
$\Gamma=0.05$ and $T=0.05$ is what we used for all the results in
the main text. For ``Bath 1'', $\Gamma=0.025$ and $T=0.05$, and
for ``Bath 2'', $\Gamma=0.05$ and $T=0.025$. }
\end{figure}

Let us also comment on the free-fermion bath coupled to the lattice.
For a flat density of states, the bath is described by two parameters:
the dissipation $\Gamma$ and the temperature $T$. In principle, one
can tune the parameters of the bath to change the final NESS. For
the simple bath used here, we can show that the effect of the bath is
not very pronounced. In Fig.~\ref{fig:bath}, we show spectral functions
for the driving frequency $\Omega=7$ and $\alpha=\frac{1}{6}$ with
different bath parameters. For $U=0$, there is a clear difference
because $\mathrm{i}\Gamma$ affects the imaginary part of $\left(G_{0}^{R\left(A\right)}\right)_{\bm{R}\bm{R}^{\prime},mn}^{-1}\left(\omega\right)$
directly. For finite $U$, there is only a minor difference. 

\bibliographystyle{apsrev4-1}

\end{document}